\documentclass[journal=jpcafh,manuscript=article,layout=twocolumn]{achemso}

\usepackage{multicol}
\usepackage{dcolumn}
\usepackage[flushleft]{threeparttable}
\usepackage[version=4]{mhchem}
\usepackage{mciteplus}
\usepackage{natmove}
\usepackage{mathrsfs} 

\makeatletter
\let\l@addto@macro\relax
\makeatother
\usepackage[fontsize=11pt]{scrextend}

\DeclareSymbolFont{matha}{OML}{txmi}{m}{it}

\title{The Sub-millimeter Rotational Spectrum of Ethylene Glycol up to 890~GHz and Application to ALMA Band 10 Spectral Line Data of NGC 6334I }

\author{Mattia Melosso}
\affiliation[ciamician]{Dipartimento di Chimica ``Giacomo Ciamician",
           Universit\`{a} di Bologna, Via Selmi 2, I-40126
           Bologna, Italy}
\email{mattia.melosso2@unibo.it}
           
\author{Luca Dore}
\affiliation[ciamician]{Dipartimento di Chimica ``Giacomo Ciamician",
           Universit\`{a} di Bologna, Via Selmi 2, I-40126
           Bologna, Italy}
           
\author{Filippo Tamassia}
\affiliation[toso]{Dipartimento di Chimica Industriale ``Toso Montanari",
           Universit\`{a} di Bologna, Viale del Risorgimento 4, I-40136
           Bologna, Italy}
           
\author{Crystal L. Brogan}
\affiliation[observatory]{National Radio Astronomy Observatory, Charlottesville, VA 22903, USA}

\author{Todd R. Hunter}
\affiliation[observatory]{National Radio Astronomy Observatory, Charlottesville, VA 22903, USA}
           
\author{Brett A. McGuire}
\affiliation[observatory]{National Radio Astronomy Observatory, Charlottesville, VA 22903, USA}
\alsoaffiliation[harvard]{Center for Astrophysics $|$ Harvard \& Smithsonian, Cambridge, MA 02138, USA}           
\email{bmcguire@nrao.edu}


\begin{document}

\maketitle

\begin{abstract}
    The rotational spectrum of the most stable conformer of ethylene glycol (\ce{HO(CH2)2OH}) has been recorded between 360--890\,GHz using a frequency-modulation sub-millimeter spectrometer.  The refinement and extension of the spectroscopic parameters over previous efforts provides predicted catalog frequencies for ethylene glycol with sufficient accuracy for comparison to high-frequency astronomical data.  The improvement in the cataloged line positions, and the need for improved accuracy enabled by high-frequency laboratory work, is demonstrated by an analysis of ethylene glycol emission at 890\,GHz in the high-mass star-forming region NGC 6334I in ALMA Band 10 observations. The need for accurate rotational spectra at sub-millimeter wavelengths/THz frequencies is discussed.
\end{abstract}

\section{Introduction}

Of the more than 200 molecules detected in the interstellar medium (ISM) to date, more than 80\% were discovered through observations of their pure rotational transitions \cite{McGuire:2018mc}.  With few exceptions\cite{Woods:1975vi,Cernicharo:2008wi,Pety:2012cp}, the detection and characterization of molecules in the ISM is preceded by high-resolution spectroscopic investigations in the laboratory.  Depending on the molecular structure, many thousands of line frequencies may be measured and fit using common software packages\cite{Pickett:1991cv,Drouin:2017ol} to a Hamiltonian\cite{Watson:1977vk}, providing a set of rotational constants ($A$, $B$, $C$) as well as higher-order terms that account for the effects of centrifugal distortion, internal motion, fine and hyperfine coupling, etc.  These constants can then be used together with the Hamiltonian to predict energy levels and associated transition frequencies not directly measured within the laboratory. The accuracy of these predictions outside the range of the laboratory measurements varies substantially depending on the structure of the molecule and the extent of the extrapolation\cite{Carroll:2010gt,melosso2017terahertz}.

Substantial laboratory efforts were undertaken about a decade ago to enable analysis of high-resolution spectra from the Heterodyne Instrument for the Far Infrared\cite{deGraauw:2010gy} (HIFI) instrument on the \emph{Herschel Space Observatory}\cite{Pilbratt:2010en}, which operated from 480--1910\,GHz. While transformative, the observations obtained with \emph{Herschel}/HIFI primarily contained spectra of either small molecules such as \ce{HCl+}\cite{DeLuca:2012cv,Gupta:2012ga} or of bright, abundant species such as \ce{CH3OH}\cite{Crockett:2014er,Neill:2014cb,Xu:2008gs}.  In general, high-frequency laboratory spectra of larger, more complex species have not been a priority, as the observational tools needed to observe them at high frequencies have not been available.

Recent observations in Band 10 (787 -- 950\,GHz) with the Atacama Large Millimeter/sub-millimeter Array (ALMA) have demonstrated that these priorities must now shift to larger molecules.  The small single-dish antennas on \emph{Herschel} or the Stratospheric Observatory for Infrared Astronomy (SOFIA)\cite{Temi:2014cb} result in large beams on the sky, heavily diluting the signal from compact star-forming regions such that (in general) the spectra are dominated by bright, abundant species like \ce{CH3OH}.  Interferometric observations with ALMA, however, have beams well-matched to the molecular emission, and are producing line-confusion limited spectra that are rich with emission from large, complex species that are not yet well studied in the laboratory.\cite{McGuire:2018bz}  

One such species is ethylene glycol (EG; \ce{HO(CH2)2OH}); prior laboratory work on the most stable conformer \textit{aGg'} of this species included spectra up to 370\,GHz\cite{christen1995rotational,christen2003millimeter}.  Here, we extend this work to 890\,GHz and show that not only is EG present and bright in our ALMA Band~10 spectra, but that identification of these lines would not be possible via extrapolation from the lower-frequency work.

\section{Experimental Section}
\subsection{Laboratory Spectroscopy and Analysis}

\begin{table*}[t]
        \caption{\label{t1}Spectroscopic parameters determined for \emph{aGg'} EG.}
        \centering
        \scalebox{0.85}{
        \begin{threeparttable}
        \begin{tabular}{p{1cm}l|D{.}{.}{10}D{.}{.}{10}|p{1cm}l|D{.}{.}{10}D{.}{.}{10}}
                \hline\hline
                \noalign{\smallskip}
                \multicolumn{2}{l}{Parameter} & \multicolumn{1}{c}{Present work} & \multicolumn{1}{c}{Previous$^{[a]}$} &
                \multicolumn{2}{l}{Parameter} & \multicolumn{1}{c}{Present work} & \multicolumn{1}{c}{Previous$^{[a]}$} \\
                \hline
                \noalign{\smallskip}
                $     A   $ &                  & 15361.18508(25) & 15361.18562(29) & $ E^*       $ &                  &  3478.89772(49) & 3478.89747(56) \\
                $     B   $ &                  & 5588.242432(61) & 5588.242718(74) & $ E^*_K     $ &                  &   -0.169548(16) &  -0.169681(20) \\
                $     C   $ &                  & 4614.489630(58) & 4614.489567(76) & $ E^*_J     $ &                  &  -0.1623385(41) & -0.1623042(63) \\
                $   D_{J} $ & $\times 10^{3}$  &    7.365648(69) &      7.5593(12) & $ E^*_{KK}  $ & $\times 10^{3}$  &   -0.036267(76) &    -0.1178(16) \\
                $   D_{JK}$ &                  & -0.03196862(60) &  -0.0315764(75) & $ E^*_{JK}  $ & $\times 10^{3}$  &    0.013813(61) &     0.0965(16) \\
                $   D_{K} $ &                  &   0.0762076(15) &   0.0736018(17) & $ E^*_{JJ}  $ & $\times 10^{9}$  &      -4.169(15) &     -5.689(38) \\
                $   d_{1} $ & $\times 10^{3}$  &   -2.332058(37) &    -2.32800(12) & $ E^*_{JJK} $ & $\times 10^{9}$  &      -0.980(52) &                \\
                $   d_{2} $ & $\times 10^{3}$  &   -0.175901(53) &    -0.16925(13) & $ E^*_{2}   $ &                  &  -0.0204125(28) & -0.0204279(39) \\
                $   H_{J} $ & $\times 10^{6}$  &   -0.012345(19) &   -0.011136(45) & $ E^*_{2J}  $ & $\times 10^{6}$  &     -0.9083(34) &    -0.9255(36) \\
                $  H_{JK} $ & $\times 10^{6}$  &     0.29653(41) &     0.32928(99) & $ E^*_{2K}  $ & $\times 10^{3}$  &    -0.01254(33) &                \\
                $  H_{KJ} $ & $\times 10^{6}$  &     -1.7613(19) &     -1.8174(21) & $ E^*_{2KJ} $ & $\times 10^{9}$  &      -2.925(51) &                \\
                $   H_{K} $ & $\times 10^{6}$  &      2.9962(34) &      3.0003(39) & $ E^*_{2JJ} $ & $\times 10^{12}$ &       -9.35(85) &                \\
                $   h_{1} $ & $\times 10^{9}$  &      -4.338(12) &      -3.611(19) & $ E^*_{4}   $ & $\times 10^{6}$  &     -0.1020(76) &      0.623(19) \\
                $   h_{2} $ & $\times 10^{9}$  &     -0.0990(90) &       0.318(18) & $ E^*_{4J}  $ & $\times 10^{9}$  &    -0.02659(77) &                \\
                $   h_{3} $ & $\times 10^{9}$  &      0.1005(38) &                 & $ F_{bc}    $ &                  &   -30.40354(16) &  -30.39749(27) \\
                $   L_{J} $ & $\times 10^{12}$ &      0.3077(25) &                 & $ F_{bc,K}  $ &                  &   -0.023612(71) &   -0.02977(19) \\
                $ L_{JJK} $ & $\times 10^{12}$ &      -3.694(57) &                 & $ F_{bc,J}  $ & $\times 10^{3}$  &     0.76356(77) &     0.7526(10) \\
                $  L_{JK} $ & $\times 10^{12}$ &        3.74(84) &                 & $ F_{bc,KK} $ & $\times 10^{6}$  &       -1.89(20) &      -4.59(20) \\
                $ L_{KKJ} $ & $\times 10^{9}$  &      0.0578(21) &                 & $ F_{bc,JK} $ & $\times 10^{6}$  &       0.249(13) &     -0.766(41) \\
                $   L_{K} $ & $\times 10^{9}$  &     -0.1200(21) &                 & $ F_{2bc}   $ & $\times 10^{3}$  &     -0.2899(11) &    -0.3686(25) \\
                $   l_{1} $ & $\times 10^{12}$ &      0.1557(17) &                 & $ F_{2bc,J} $ & $\times 10^{9}$  &        6.79(13) &                \\
                $   l_{2} $ & $\times 10^{12}$ &      0.0186(13) &                 & $ F_{ab}    $ &                  &  -143.08505(38) & -143.08745(52) \\
                $   l_{3} $ & $\times 10^{12}$ &     0.01645(84) &                 & $ F_{ab,J}  $ & $\times 10^{3}$  &      1.9355(12) &     2.1674(41) \\
                $   l_{4} $ & $\times 10^{15}$ &        1.00(42) &                 & $ F_{ab,JK} $ & $\times 10^{6}$  &     -0.1883(96) &     -0.471(14) \\
                            &                  &                 &                 & $ F_{ab,K}  $ & $\times 10^{3}$  &                 &       7.83(16) \\
                            &                  &                 &                 & $ F_{2ab}   $ & $\times 10^{3}$  &     -1.4641(27) &     -0.821(12) \\
                            &                  &                 &                 & $ F_{2ab,J} $ & $\times 10^{6}$  &    -0.03102(79) &                \\
                \hline
                \noalign{\smallskip}
                Lines           &                 &  \multicolumn{1}{c}{777}           &  \multicolumn{1}{c}{}      \\
                rms             &                 &  \multicolumn{1}{c}{0.064}         &  \multicolumn{1}{c}{}    \\
                $\sigma$        &                 &  \multicolumn{1}{c}{0.86}          &  \multicolumn{1}{c}{}     \\
                \hline
        \end{tabular}
        
        \begin{tablenotes}
        \small
	    \item \textbf{Notes:} Units are in MHz, except the dimensionless standard deviation $\sigma$. Values in parenthesis denote one standard deviation and apply to the last digits of the constants. $[a]$ \citet{christen2003millimeter}.
	    \end{tablenotes}
		\end{threeparttable}
		}
\end{table*}

The rotational spectrum of the most stable conformer of EG has been recorded with a frequency-modulation sub-millimeter spectrometer \cite{melosso2019astronomical}. Spectral coverage between 360 and 890\,GHz was attained by combination of a Gunn diode (J.E. Carlstrom Co\cite{Carlstrom:1985vy}) emitting in the range 80--115\,GHz and a number of passive multipliers (VDI, RPG)\cite{Crowe:2007kj} in cascade. The resulting output radiation power was \emph{ca.} 1\,mW, 0.1\,mW, and 10\,$\mu$W at 360--390, 520--620, and 780--890\,GHz, respectively. The Gunn's radiation was phase-locked to a harmonic of a frequency synthesizer (Schomandl, ND 1000) and frequency modulated at $f$ = 48\,kHz by the 75\,MHz reference signal. A free-space glass absorption cell (3.25\,m long, 5\,cm in diameter) filled with EG vapors at a static pressure of \emph{ca.} 15\,mTorr was used for the measurements. Phase-sensitive detection at twice the modulation frequency was employed, so that the $2f$ spectrum profile was observed. A liquid-helium-cooled InSb bolometer (QMC Instr. Ltd. type QFI/2) was used as detector. We estimate the measurement uncertainty on any given line to between 25 and 50\,kHz, depending on the linewidth and the signal-to-noise ratio (SNR) of the spectrum at that frequency.

\begin{figure*}[htb!]
    \centering
    \includegraphics[width=0.9\textwidth]{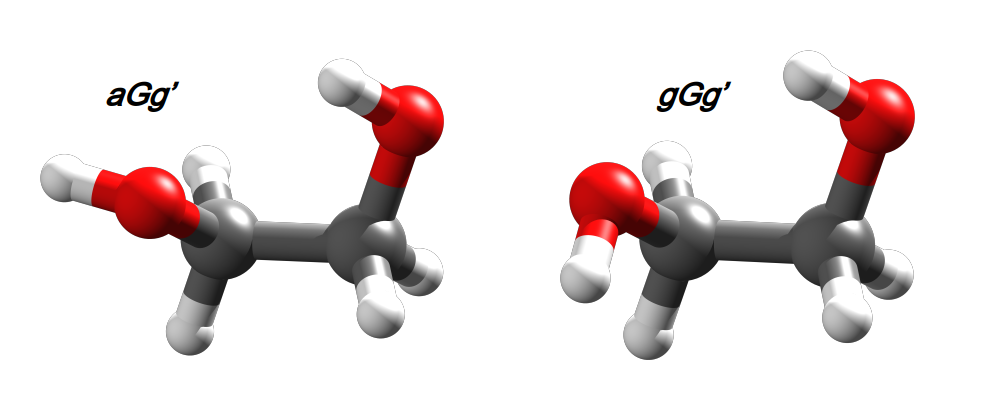}
    \caption{The \textit{aGg'} and \textit{gGg'} conformers of EG. The capital $G$ indicates the \textit{gauche} arrangement of the OH groups with respect to rotation at the \ce{C-C} bond, while lowercase letters \textit{a} (\textit{anti}) or \textit{g} indicate the position of hydroxyl hydrogen atoms with respect to rotation at the \ce{C-O} bonds. Clockwise and counterclockwise rotations are indicated by the symbols \textit{g} and \textit{g'}, respectively. The two isomeric forms are separated by a rotational barrier of 740\,K\cite{christen2001rotational}.}
    \label{fig:conf}
\end{figure*}

From a spectroscopic point of view, the \textit{aGg'} conformer of EG (shown in Fig.~\ref{fig:conf}) is a nearly-prolate asymmetric-top rotor with an internal large amplitude motion. Because of the tunneling between two equivalent positions of the hydroxy groups, each rotational energy level $J_{Ka,\,Kc}$ is split into two sub-levels, labeled as 0 or 1. The state 1 is slightly higher in energy than the state 0.
EG possesses a permanent electric dipole moment of 2.33 D, with components $\mu_a=2.080(3)$ D, $\mu_b=0.936(7)$ D, and $\mu_c=-0.47(1)$ D\cite{christen1995rotational}. Since the inversion motion causes the change in sign of $\mu_a$ and $\mu_c$, these two components allow transitions between the tunneling states 0 and 1; on the other hand, $b$-type transitions take place within the states.

The rotational-torsional energy levels can be described using the formalism introduced by \citet{christen2003millimeter}.
Rather than fitting two sets of spectroscopic constants (one for each tunneling state) along with the energy difference between levels 0 and 1, as is often performed for these types of analyses, \citet{christen2003millimeter} have rearranged the Hamiltonian so that a unique set of rotational constants and the energy difference are fit along with their centrifugal dependencies.
The latter (Reduced Axis System; RAS) and former (Internal Axis Method; IAM) approaches have been demonstrated to be equivalent; the use of RAS method has been preferred in this and previous works\cite{christen2003millimeter,muller2004millimeter} because it is implemented in the SPFIT program\cite{Pickett:1991cv,Pickett:1997kz}.
An extensive description of the RAS Hamiltonian is given in the original papers\cite{christen2003millimeter,muller2004millimeter}.

Based on the spectroscopic parameters reported previously\cite{christen2003millimeter}, the rotational spectrum of \textit{aGg'}-EG were predicted and investigated at frequencies above 360\,GHz. Mostly, rotational-torsional transitions have been recorded line-by-line, scanning a few MHz around the predicted position. Spectral predictions were continuously updated by refining the spectroscopic constants on the basis of our newly recorded transitions, in order to avoid possible misassignment. Indeed, discrepancies up to 36\,MHz were found between older and actual line frequencies.  The observed lineshapes were dominated by Doppler broadening; the pressure of EG vapors was maintained at a value which allowed the observation of spectra with good SNR without causing significant pressure broadening. The uncertainty in line center frequencies was evaluated on a line by line basis, accounting for the effects of Doppler broadening on the linewidths.

In total, 228 distinct frequencies\footnote{The list of newly observed transitions, along with their residuals from the final fit, is provided as supplementary material.}, corresponding to 370 transitions, have been added to the existing dataset and all the experimental data have been fitted together in a least-square procedure in which each datum was weighted accordingly to its uncertainty. We probed rotational energy levels up to $J' = 86$ and $K_a' = 29$, thanks to which we were able to substantially improve the centrifugal analysis of \textit{aGg'}-EG. For instance, the whole set of centrifugal distortion terms up to the 8th order of the rotational Hamiltonian ($L$ and $l$ constants) has been determined for the first time with good accuracy. Some centrifugal dependencies of Coriolis ($F_{bc}$ and $F_{ab}$) and energy difference ($E^*$) terms have also been included in the analysis. To limit strong correlations among the spectroscopic parameters, however, we tried to keep the number of floating constants low, as long as the standard deviation of the fit remained at values around 1. On the whole, our newly determined parameters agree well with those previously reported\cite{christen2003millimeter} but their associated errors are slightly smaller in our work, as can be seen from Table~\ref{t1}.
The overall standard deviation ($\sigma = 0.86$) seems to indicate that the analyzed data are well-reproduced within their expected accuracy. Our new spectral predictions offer a solid base to assist astronomical searches of \textit{aGg'}-EG even at high frequencies.

Regrettably, a thorough knowledge of the rotational spectrum of EG is still a distant goal. This is because (i) the spectrum of the second most stable conformer \textit{gGg'} (which is estimated to be only 160--290\,K more energetic than \textit{aGg'}\cite{kristiansen1987microwave,christen2001rotational,muller2004millimeter}) has been studied up to 579\,GHz but extrapolations to higher frequencies were suggested to be not reliable \cite{muller2004millimeter} and (ii) transitions belonging to low-lying vibrational excited states remain completely unassigned.  The potentially substantial impact of such vibrational states on the crowding of astronomical spectra has previously been demonstrated \cite{Fortman:2012is}.

\subsection{Observations and Analysis}

The observations presented here are from ALMA projects \#2017.1.00717.S, \#2017.1.00661.S, and \#2015.A.00022.T.  The calibration and reduction are presented in detail elsewhere,\cite{McGuire:2017gy,Hunter:2017th,Brogan:2018wb,McGuire:2018bz} and will only be briefly discussed here.  The observations were centered at $\alpha$(J2000) = 17:20:53.36, $\delta$(J2000) = -35:47:00.0, and have been smoothed to provide a uniform angular resolution of $0.26''\times 0.26''$ across all datasets.  More than a dozen portions of frequency coverage span the range from \mbox{130\,--\,898\,GHz} with velocity resolutions of \mbox{$\sim$0.3\,--\,1.0}\,km\,s$^{-1}$.  The continuum has been identified from (relatively) line-free channels and subtracted from the spectra following the methods described in \citet{Brogan:2018wb}.

We have simulated the spectrum of EG up to 900\,GHz using both the new spectroscopic parameters derived here, and the literature values from which current database entries are derived.\cite{christen2003millimeter}  The spectra are simulated following the formalisms of \citet{Turner:1991um}, which include corrections for optical depth based on the background continuum temperatures we have measured during the continuum subtraction.  Intensities were converted to brightness temperature on the Planck scale, as errors introduced from the Rayleigh-Jeans approximation become large at high frequencies and small angular scales.  

\begin{figure*}[htb!]
\centering
\includegraphics[width=\textwidth]{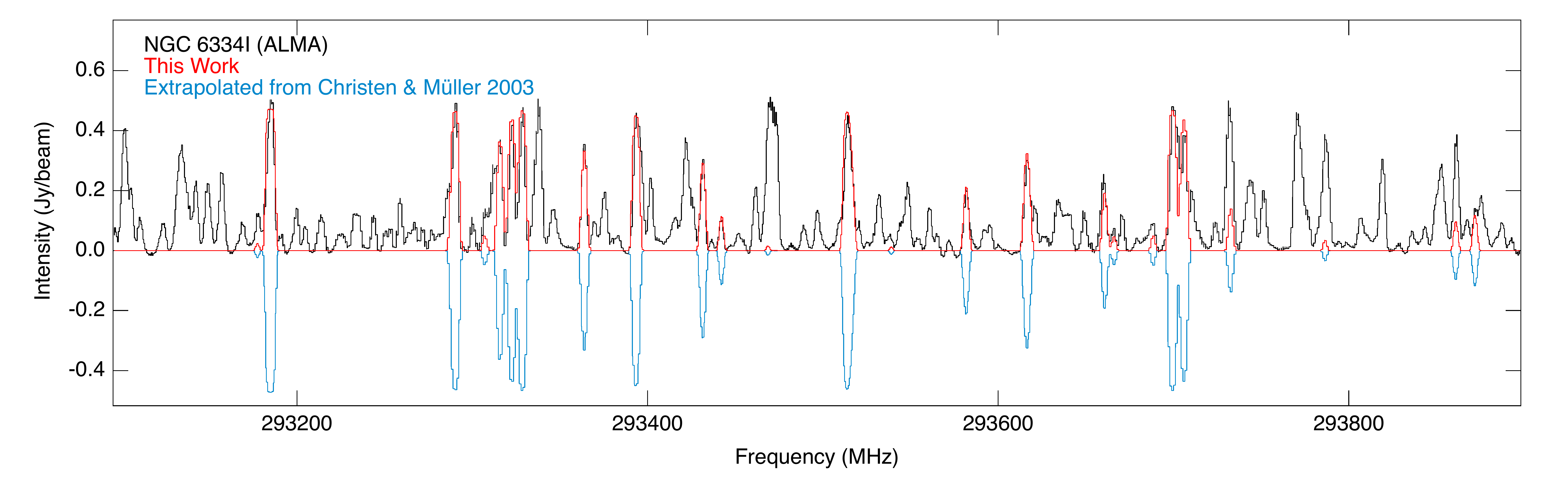}
\includegraphics[width=\textwidth]{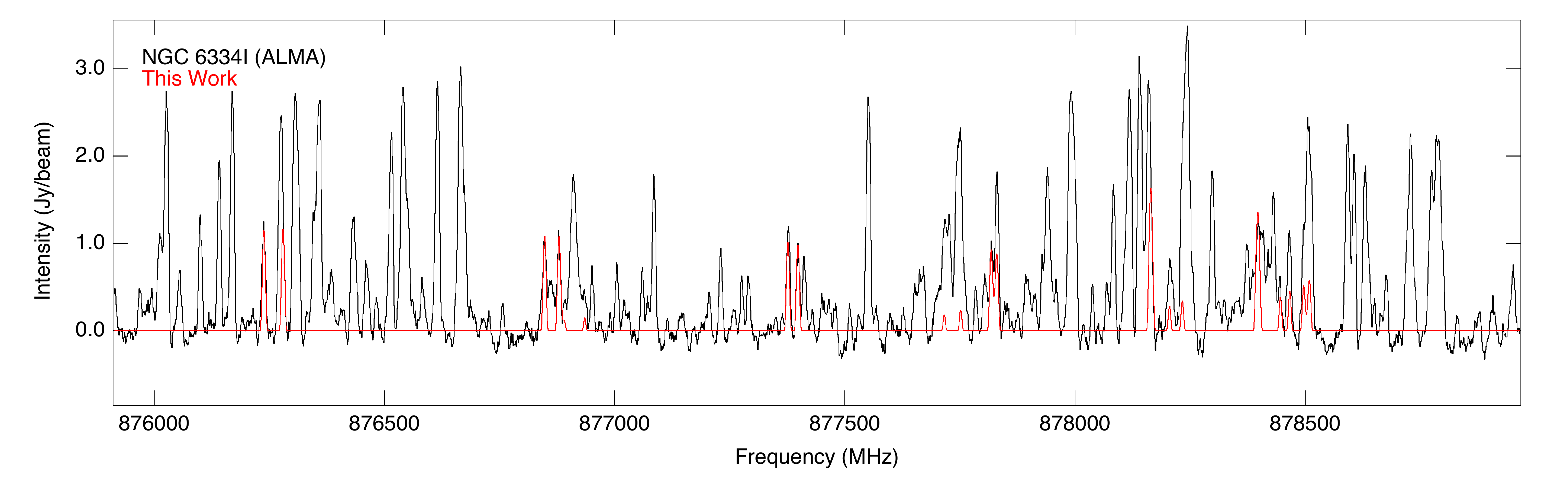}
\includegraphics[width=\textwidth]{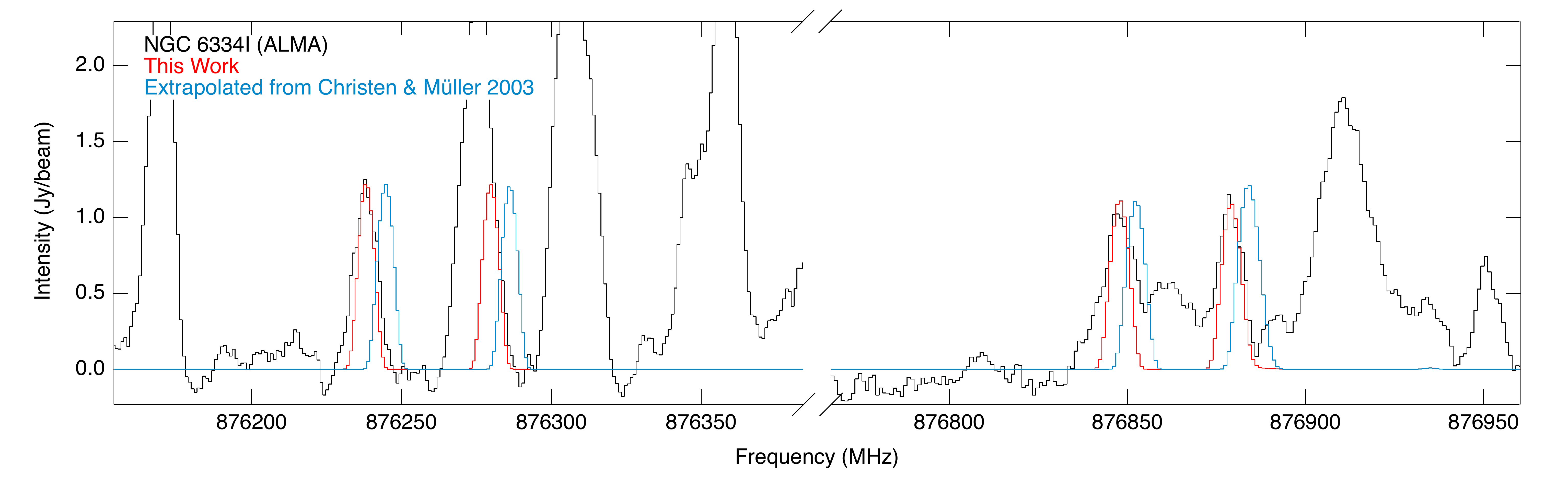}
\caption{\textbf{(Top)} ALMA spectra of NGC 6334I in Band 7, with simulated lines of EG from the analysis carried out in this work overlaid in red and using the constants derived in \citet{christen2003millimeter} inverted and in blue.  The simulation was performed at $T_{ex}$~=~135\,K, $\Delta V$~=~3.2\,km\,s$^{-1}$, and $v_{lsr}$~=~-7\,km\,s$^{-1}$.  There is essentially perfect frequency agreement between the datasets. \textbf{(Middle)} ALMA spectra of NGC 6334I in Band 10, again with simulations from this work in red.  \textbf{(Bottom)} ALMA spectra of NGC 6334I in Band 10, again with simulations from this work in red and using the constants derived in \citet{christen2003millimeter} inverted and in blue, zoomed in to show the frequency disagreement.}
\label{alma}
\end{figure*}

For complex molecules such as glycolaldehyde (\ce{HOCH2CHO}) with optically thin emission, we have found that a single column density with an excitation temperature of $T_{ex}$~=~135\,K, linewidth of $\Delta V$~=~3.2\,km\,s$^{-1}$, and $v_{lsr}$~=~-7\,km\,s$^{-1}$ well reproduces the observed spectral profiles.\cite{McGuire:2018bz}  For optically thick molecules, such as \ce{CH3OH} and EG however, the excitation temperature and column density can vary more significantly across the ALMA Bands due to varying penetration depths of the observations into the cloud.\cite{McGuire:2018bz} A complete treatment of these species is not only beyond the scope of this work, but would require a multidimensional radiative transfer model due to the spatial complexity of the region.  Therefore, here we limit ourselves to only discussing the impact that the frequency accuracy of the predictions has on line identification in high frequency ALMA data.  A comparison of EG emission simulated using the existing literature constants to those derived in this work is shown in Figure~\ref{alma}.

\section{Discussion}

In Band 7 (Fig.~\ref{alma} Top), well within the frequency range of the previous laboratory work, there is, as expected, near perfect agreement between not only spectra simulated with our refined constants compared to those of \citet{christen2003millimeter}, but also between both simulations and the observations.  The middle panel shows ALMA Band 10 data with only the newly measured lines overlaid.  Both the top and middle panels demonstrate the need for robust line catalogs, as even relatively complex species such as EG contribute substantially to the total line density in ALMA observations of hot cores.  

The bottom panel demonstrates the need for high frequency laboratory work.  The simulation based on the prior work is significantly offset from the true line positions, as shown by the excellent agreement between our new predictions and the observations.  This would result either in a faulty conclusion that EG is not present in the Band 10 spectra, or perhaps that it is velocity shifted to an incorrect $v_{lsr}$.  

\begin{figure}
    \centering
    \includegraphics[width=\columnwidth]{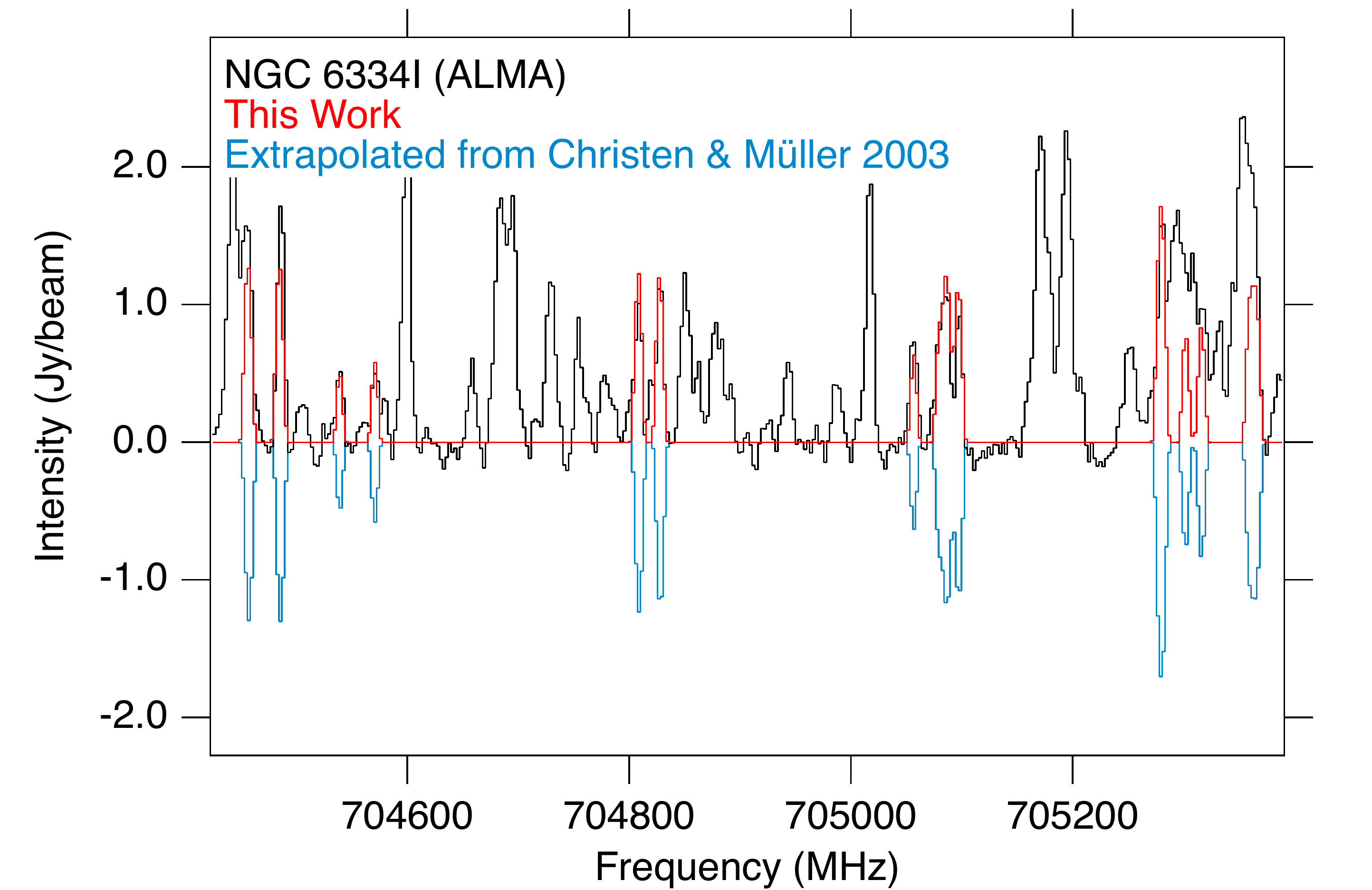}
    \caption{ ALMA spectra of NGC 6334I in Band 9, with simulated lines of EG from the analysis carried out in this work overlaid in red and using the constants derived in \citet{christen2003millimeter} inverted and in blue.  The simulation was perfomed at $T_{ex}$~=~135\,K, $\Delta V$~=~3.2\,km\,s$^{-1}$, and $v_{lsr}$~=~-7\,km\,s$^{-1}$.  There is essentially perfect frequency agreement between the datasets.}
    \label{band9}
\end{figure}

Many of the lines in our Band 10 spectra are shifted by $\sim$5\,MHz (2\,km\,s$^{-1}$ at 890\,GHz).  Yet, notably, stepping down in frequency to observations in ALMA Band 9 (602--720\,GHz), the two datasets are once again in excellent agreement with each other and with the observations (Fig.~\ref{band9}).  Indeed, because extrapolation errors are dependent on the quantum numbers and the accuracy with which the corresponding energy levels are known, rather than directly upon frequency, the situation can be quite challenging to assess.  For example, higher-frequency transitions of one type, e.g. \emph{b}-type transitions, arising between two energy levels that are well-characterized from lower-frequency measurements of \emph{a}-type transitions may be accurately predicted.  On the other hand, transitions very near in frequency but arising from poorly constrained energy levels might be highly inaccurate.  Additionally,  uncertainties in higher-order distortion terms scale non-linearly with increasing values of the quantum numbers, 

As a result, it is often not clear the extent to which extrapolations become dangerous.  Extrapolating a factor of 1.9 times higher in frequency (to 700\,GHz) than the original work by \citet{christen2003millimeter} results in predictions that can be used to accurately analyze observational data, because these energy levels are well-constrained by the laboratory measurements.  A factor of 2.4 (to 875\,GHz), however, is too far, as demonstrated by our Band 10 analysis. 

Most spectral fitting programs, including SPFIT/SPCAT, provide estimated uncertainties on the line positions that could be used for a zeroth-order approximation of the safety of extrapolation.  Yet, these are often statistical errors derived from the fit itself, and are not representative of the overall uncertainty of the prediction. For example, the EG signal just below 876250\,MHz in Fig.~\ref{alma} arises from the $29_{29,0} - 28_{28,1}$ and $29_{29,1} - 28_{28,0}$ transitions.  Using the errors in the constants reported by \citet{christen2003millimeter}, SPCAT provides an uncertainty of $\sim$0.3\,MHz for these line centers, a factor of $\sim$15 lower than the actual error.  The most reliable and robust method for ensuring that sufficiently accurate data are available is to measure these species in the laboratory at the required frequencies.

\section{Conclusions}

Here, we have extended the investigation of the rotational spectrum of the lowest-energy conformer (\textit{aGg'}) of ethylene glycol to 890\,GHz.  The analysis and assignment was significantly aided by prior lower-frequency work.  Extrapolating frequencies based on that lower-frequency work proved robust for analyzing astronomical spectra from ALMA toward the high-mass star-forming region NGC 6334I up through $\sim$800\,GHz.  In ALMA Band 10, however, these predictions diverged substantially, requiring our new laboratory measurements to make a secure detection.  This demonstrates the need for new high-frequency laboratory experiments to enable the robust analysis of data in corresponding ALMA observations.

\section{Associated Content}

\textbf{Supporting Information.}  List of newly observed transitions and their residuals from the final fit, SPCAT input and output files
(aGg.int, aGg.var, aGg.cat).

\section{Acknowledgements}

The authors thank the anonymous reviewers for helpful comments which have improved the quality of this manuscript.  This work has been supported in Bologna by MIUR `PRIN 2015' funds (project ``STARS in the CAOS (Simulation Tools for Astrochemical Reactivity and Spectroscopy in the Cyberinfrastructure for Astrochemical Organic Species)'' - Grant Number 2015F59J3R) and by the University of Bologna (RFO funds). This paper makes use of the following ALMA data: ADS/JAO.ALMA\#2017.1.00717.S, \#2017.1.00661.S, and \#2015.A.00022.T. ALMA is a partnership of ESO (representing its member states), NSF (USA) and NINS (Japan), together with NRC (Canada) and NSC and ASIAA (Taiwan) and KASI (Republic of Korea), in cooperation with the Republic of Chile. The Joint ALMA Observatory is operated by ESO, AUI/NRAO and NAOJ.   The National Radio Astronomy Observatory is a facility of the National Science Foundation operated under cooperative agreement by Associated Universities, Inc.  Support for B.A.M. was provided by NASA through Hubble Fellowship grant \#HST-HF2-51396 awarded by the Space Telescope Science Institute, which is operated by the Association of Universities for Research in Astronomy, Inc., for NASA, under contract NAS5-26555.  This research made use of NASA’s Astrophysics Data System  Bibliographic  Services,  Astropy,  a community-developed core Python package for Astronomy \citep{astropy}, and APLpy, an open-source plotting package for Python hosted at http://aplpy.github.com.

\section{Table of Contents Graphic}

\begin{figure}
    \centering
    \includegraphics[width=3.25in]{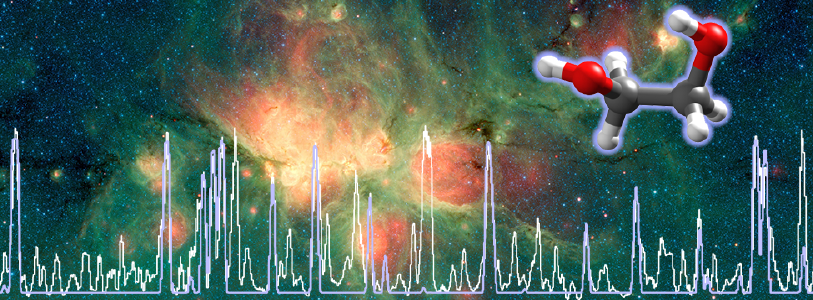}
    \caption{Table of Contents Graphic}
\end{figure}

\bibliography{EG,bibliography}

\end{document}